\documentstyle[12pt]{article}

\begin{document}

\author{Ram\'on Risco-Delgado}
\title{Bell's inequalities and indeterminism}
\date{September, 2001 }
\maketitle

\begin{abstract}
The inequality of Clauser and Horne [{\em Phys. Rev.} {\em D} {\bf 10}, 526
(1974)], intended to overcome the limited scope of other inequalities to
deterministic theories, is shown to have a resticted validity even in case of perfect detectors and perfect angular correlations.
\end{abstract}

In the current debate on local hidden variables (l.h.v.) \cite{Rowe} we try
to clarify some points around one of the inequalities that there exists in
the literature in order to contrast quantum mechanics with local realism.

Such inequalities begin with Bell's \cite{bell}. He proved that, if we do an
experiment in which two spatially separated measurements are performed with
perfect detectors, the correlations of their results must satisfy a certain
inequality if they are to be explained by a local deterministic theory.
There are two conditions for this inequality, namely determinism and perfect
detectors. In order to perform an experiment with optical photons, for which
the effective efficiency is still nowadays rather low, Clauser, Horne,
Shimony and Holt \cite{CHSH} derived an inequality from the assumptions of
local realism, determinism and, what they called the ``fair sampling
assumption". This inequality could be carried to the laboratory and it was
found to be experimentally violated. However there was a class of local
realistic theories, those with a essential stochasticity (or
nondeterministic) that were not tested.

The next step was made by Clauser and Horne (CH) \cite{CH}. Their aim was to
obtain an inequality for low efficiency, valid for what they called {\it %
Local Objective Theories}, which include such nondeterministic theories. The
article begins by establishing the following

\begin{quotation}
Consider the state specification of the...system at a time intermediate
between its emission and its impinging on either measuring apparatus. Denote
this state by $\lambda $. Note that we do not necessarily make a commitment
to the completeness of this state specification...As the state described
initially by $\lambda $ evolves, it may or may not trigger a count at
apparatus 1, and similarly it may or may nor do so at apparatus 2. The
initial state $\lambda $, if it serves the same role as in existing
theories, will suffice to determine {\em at least} the probabilities of
these events. Let the probabilities of a count being triggered at apparatus
1 and 2 be $p_1(\lambda ,a)$ and $\ p_2(\lambda ,b)$ respectively, and let $%
p_{12}(\lambda ,a,b)$ be the probability that both counts are triggered.
\end{quotation}

They then introduced {\em factorability} as a ``reasonable" locality
condition. In their words ``the factored form is a natural expression of a
field-theoretical point of view, which in turn is an extrapolation from the
common-sense view that there is no action at distance." The factorability
assumption is

\begin{equation}
\label{fact}{p_{12}(\lambda ,a,b)=}p_1(\lambda
,a)\ p_2(\lambda ,b)
\end{equation}
In appendix A of their article they proved that the quantity 
\begin{equation}
\label{U}
\begin{array}{c}
U(a,a^{\prime },b,b^{\prime },\lambda )=p_1(\lambda ,a)\ p_2(\lambda
,b)-p_1(\lambda ,a)\ p_2(\lambda ,b^{\prime })+p_1(\lambda ,a^{\prime })\
p_2(\lambda ,b) \\ 
+p_1(\lambda ,a^{\prime })\ p_2(\lambda ,b^{\prime })-p_1(\lambda ,a^{\prime
})-\ p_2(\lambda ,b)
\end{array}
\end{equation}
satisfies $-1\leq U(a,a^{\prime },b,b^{\prime },\lambda )\leq 0$. By
replacing, in this inequality, the product of probabilities by the joint
probability (from (\ref{fact})), integrating over $\rho (\lambda )\ d\lambda $,
and with the aid of the ``no-enhancement'' hypothesis [$p_1(\lambda ,a)\leq
p_1(\lambda ,\infty )$, where $p_1(\lambda ,\infty )$ is the detection
probability with the polarizer removed, and similarly for $p_2(\lambda ,b)$%
], the CH inequality follows: 
\begin{equation}
\label{CH}-p_{12}(\infty ,\infty )\leq p_{12}(a,b)-p_{12}(a,b^{\prime
})+p_{12}(a^{\prime },b)+p_{12}(a^{\prime },b^{\prime })-p_{12}(a^{\prime
},\infty )-p_{12}(\infty ,b)\leq 0
\end{equation}

We shall show that i) there exist local realistic situations that do not
satisfy (\ref{fact}), and ii) that those situations in which it is satisfied
are trivial cases in which either $U(a,a^{\prime },b,b^{\prime },\lambda
)=-1 $ or $U(a,a^{\prime },b,b^{\prime },\lambda )=0$ (something both
stronger and simpler to prove than the inequality derived the appendix A of
CH).

i) Suppose that two classical systems (i.e. two particles) correlated in the source are
emitted toward two opposite detectors. The particles can be in one of two
states to be measured, that we shall call ``up'' and ``down''. The
correlation is such that the quantity to be measured has equal values for
the two systems, that is, there are only cases of ``up-up'' or ``down-down''
of emissions from the source, each with, let's say $1/2$ of essential,
irreducible probability. Of course the systems can have l.h.v. $\lambda $,
but $\lambda $ does not suffice to fix the results of the measurement. For a
given value of $\lambda $ we have $p_1(\lambda ,a\equiv up)=1/2,p_2(\lambda
,b\equiv up)=1/2$, but ${p_{12}(\lambda ,a\equiv
up,b\equiv up)=1/2\neq }p_1(\lambda ,a\equiv up)\ p_2(\lambda ,b\equiv
up)=1/4$. This simple counterexample shows that (\ref{fact}) is not general
enough to embrace all forms of local realism. Similar results are obtained
for any type of correlation in the source (i.e.``up-down'' ).

ii) In case of absolute determinism $p_1(\lambda ,a)=1$ or $0$, and $%
p_2(\lambda ,b)=1$ or $0$. This means that there are sixteen cases for $U$,
as a function of the values of $p_1$and $\ p_2$. Half of them give $U=-1,$
the others give $U=0$.

To sum up, in case of {\em determinism} not only is $-1\leq U(a,a^{\prime
},b,b^{\prime },\lambda )\leq 0$ satisfied, but always either $U=-1$ or $U=0$%
. In case of {\em nondeterminism} (\ref{fact}) simply does not hold.
Therefore the question of nondeterministic local realism is left open in CH.
However we shall see that nevertheless it can be closed in the following way.

Suppose that Nature is essentially stochastic and therefore$\ p_1(\lambda
,a) $ and $p_2(\lambda ,b)$ are in general different from $0$ and $1$.Then,
there always exists another two functions $\widetilde{p}_1(\lambda ,\mu ,a)$
and $\widetilde{p}_2(\lambda ,\mu ,b)$, which take only values $0$ or $1$,
defined in the following way: $\widetilde{p}_1(\lambda ,\mu ,a)=1$ if $0\leq
\mu \leq $ $p_1(\lambda ,a)$; $\widetilde{p}_1(\lambda ,\mu ,a)=0$ if $%
p_1(\lambda ,a)\leq \mu \leq $ $1$ (similar for $\widetilde{p}_2(\lambda
,\mu ,b)$). One is free to believe or not in the real existence of the
hidden variable $\mu $. The point is that because $\ p_1(\lambda ,a)=\int $ $%
\widetilde{p}_1(\lambda ,\mu ,a)\ d\mu $ (similar for $p_2(\lambda ,b)$)
then the factual quantity $\ p_1(a)$ can always be the result of the
integration over the set of hidden variables $\lambda $ and $\mu $: $\
p_1(a)=\int $ $\widetilde{p}_1(\lambda ,\mu ,a)\ d\lambda \ d\mu $ (similar
for $p_2(b)$). This result means that any essentially stochastic theory can
be included within a more general deterministic theory. Of course, it can be
applied to quantum mechanics and therefore the inequalities can be used in
principle to contrast it versus local realism. On the other hand, this
result is a prove of the room for hidden variables. The consequence for the
CH inequality is that the nondeterministic case is also contained in this
inequality, but as a particular case of the deterministic one. This result
also applies to weakest definitions of local realism \cite{selleri}.

I am grateful for a critical reading of the manuscipt by T.W. Marshall, E. Santos and A. Casado. I also aknowledge the nice comments on this subject from M. Horne at the Institute f\"ur
ExperimentalPhysik (Innsbruck) and the friendship of the people there: M. Reck, H. Weinfurter, J.W. Pan, M. Michler, G. Wehis, B. Dopfer, T. Jennewein, D. Bowmeester and A. Zeilinger.

\end{document}